\documentclass[letterpaper, 10 pt, conference]{ieeeconf}  

\IEEEoverridecommandlockouts                              
\usepackage{graphicx}
\usepackage{amsmath}
\usepackage{amssymb}
\usepackage{booktabs}
\usepackage{amsfonts}
\usepackage{bm}
\usepackage{pifont}
\usepackage{bigdelim} 
\usepackage{multirow}  
\usepackage{url}

\usepackage[colorlinks,
            linkcolor=red,
            anchorcolor=blue,
            citecolor=green]{hyperref}
            
\usepackage[table]{xcolor} 
\usepackage{colortbl}
\usepackage{flushend}

\title{\LARGE \bf
RTA-Former: Reverse Transformer Attention for Polyp Segmentation
}

\author{Zhikai Li,$^{1} $ Murong Yi,$^{2}$ Ali Uneri,$^{2}$ Sihan Niu,$^{1}$ and Craig Jones$^{1,3,4*}$
\thanks{$^{1}$Zhikai Li, Sihan Niu, and Craig Jones are with the Department of Computer Science, Johns Hopkins University, Baltimore MD, USA
        {\tt\small zli264@jhu.edu, sniu7@jhu.edu, craigj@jhu.edu}}%
\thanks{$^{2}$Murong Yi and Ali Uneri are with the Department of Biomedical Engineering, Johns Hopkins University, Baltimore MD, USA
        {\tt\small myi16@jhu.edu, ali.uneri@jhu.edu}}%
\thanks{$^{3,4}$Craig Jones is also with the Department of Radiology and Radiological Science, Johns Hopkins University, School of Medicine, Baltimore, USA, and the Malone Center for Engineering in Healthcare, Johns Hopkins University, Baltimore MD, USA
        {\tt\small craigj@jhu.edu}}%
\thanks{$^{*}$Corresponding author}
}

\begin{document}

\maketitle
\thispagestyle{empty}
\pagestyle{empty}

\begin{abstract}
Polyp segmentation is a key aspect of colorectal cancer prevention, enabling early detection and guiding subsequent treatments. Intelligent diagnostic tools, including deep learning solutions, are widely explored to streamline and potentially automate this process. However, even with many powerful network architectures, there still comes the problem of producing accurate edge segmentation. In this paper, we introduce a novel network, namely RTA-Former, that employs a transformer model as the encoder backbone and innovatively adapts Reverse Attention (RA) with a transformer stage in the decoder for enhanced edge segmentation. The results of the experiments illustrate that RTA-Former achieves state-of-the-art (SOTA) performance in five polyp segmentation datasets. The strong capability of RTA-Former holds promise in improving the accuracy of Transformer-based polyp segmentation, potentially leading to better clinical decisions and patient outcomes. Our code is publicly available on \href{https://github.com/garlicman-man/RTA-Former}{GitHub}.
\end{abstract}

\section{INTRODUCTION}
Automatic polyp segmentation plays a pivotal role in colorectal cancer prevention by facilitating early detection and providing references for further intervention. However, even with the help of modern medical imaging techniques, qualitative interpretation often requires specialized expertise that leads to prolonged diagnosis times. 

Polyp segmentation can be formulated as a pixel-wise classification task, where many deep-learning approaches are applied. One is the U-Net\cite{ronneberger2015u} segmentation network. The U-Net proposed a structure composed of an encoder and a decoder with skip connections in between. Following solutions that use such convolutional neural networks (CNNs) inherited the structure, such as UNet++\cite{zhou2018unet++}, ResUnet++\cite{jha2019resunet++}, Pranet\cite{fan2020pranet}. These networks all show promising results for medical image segmentation.

Recently, the vision transformer (ViT)\cite{dosovitskiy2020image} shows extended capabilities. By utilizing a self-attention mechanism, a patch-level approach enables the network to establish relationships between disparate regions of the image, ViT is facilitated with the capture of global dependencies. Naturally, the ViT architecture was adapted for segmentation\cite{zheng2022deep}, such as in TransUnet\cite{chen2021transunet}. The promising results inspired a ViT variation to allow the transformer to produce hierarchical features with lower computational cost, namely the Pyramid Vision Transformer (PVT)\cite{wang2021pyramid}. This also inspired many networks to introduce PVT as an alternative to CNN backbone, instead of a coordinator, such as PolypPVT\cite{dong2021polyp}. 

Although deep learning has shown promising results in semantic segmentation, for polyp segmentation tasks where the polyp tissue and the background can be very similar, precisely localizing the edges is important and challenging. 
In this paper, we introduce RTA-Former, a novel architecture that uses PVT as an encoder and employs a Hierarchical Feature Synthesizer (HFS) composed of Reverse Transformer Attention (RTA) to improve its capability for edge segmentation. Our main contributions are threefold:

\textbf{(1) Novel Network Architecture}: We introduce the RTA-Former, which utilizes hierarchical features generated by the transformer encoder. By integrating the transformer structure into the reverse attention mechanism of the decoder, the network can focus on the difficult edge regions of the image.

\textbf{(2) Flexible Backbone Network}: Our proposed decoder architecture can be combined with different sizes of PVT backbones, to meet the different needs on the balance between computation time and performance. It then provides more options in structure depending on practical tasks.

\textbf{(3) Performance Evaluation on Multiple Datasets}: We compared the performance of RTA-Former on the polyp segmentation task on five datasets, evaluating both the learning ability and generalization. The result not only indicates that RTA-Former meets the state-of-the-art performance, but also demonstrates the flexibility of the RTA-Former structure in generalization.

\section{METHOD}
\begin{figure}
  \centering
  \includegraphics[width=1.0\linewidth]{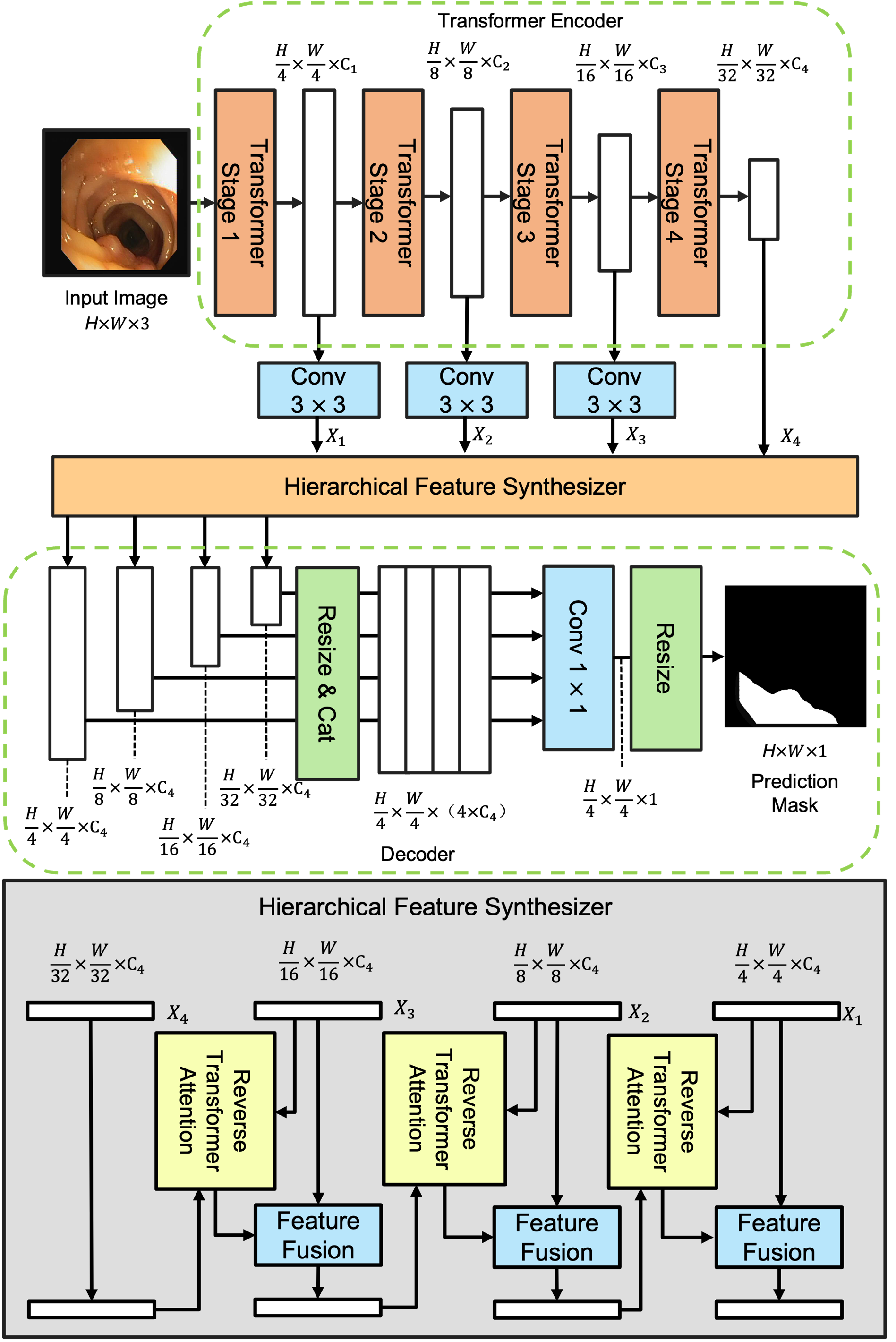}
  \caption{An overview of RTA-Former architecture. The upper section showcases the overall architecture of the RTA-Former model, which is composed of an Encoder, a Hierarchical Feature Synthesizer, and a Decoder. The lower section offers an in-depth view of the internal structure of our Hierarchical Feature Synthesizer.}
  \label{whole}
\end{figure}
For this section, we will discuss our feature fusion mechanism, then introduce our backbone structure used for the encoder, followed by an introduction to the hierarchical feature synthesizer module and its internal reverse transformer attention structure. Lastly, we dive into the architecture and functionalities of our decoder.

\subsection{Fast Feature Fusion Mechanism(FF)}
Inspired by EfficientDet\cite{tan2020efficientdet}, our Fast Feature Fusion mechanism (FF) amalgamates multi-scale features. For each feature, we assign a normalized, adaptive weight. The weighted sum of these features undergoes a Swish activation function, represented as \( f(x) = x \cdot \sigma(x) \), where \( \sigma \) is the sigmoid function. For features \( \{x_1, x_2, \ldots, x_n\} \), the fusion process is formalized as follows:
\begin{equation}
    \mathrm{FF} = \mathrm{Swish}\left(\sum_{i=1}^{n} w_i \cdot x_i\right)
\end{equation}

where \( F \) is the fused feature and \( \{w_1, w_2, \ldots, w_n\} \) are the normalized learnable weights.

\subsection{Transformer Encoder}
We adopt the transformer architecture encoder as the model's backbone to ensure sufficient universality and multi-scale feature handling capability in polyp segmentation. Specifically, we use a pre-trained PVTv2. The PVTv2 employs convolutional layers to replace traditional patch embedding modules in a transformer. This allows the continuous capture of spatial information and features at various levels. To generate the final features for input to the Hierarchical Feature Synthesizer module, the output features are passed through a \(3 \times 3\) convolutional layer for channel dimension upscaling.

\subsection{Hierarchical Feature Synthesizer (HFS)}

The Hierarchical Feature Synthesizer (HFS) is a novel architecture tailored for advanced feature extraction. Rooted in the transformer-based encoder, it assimilates hierarchical feature representations. Specifically, this module leverages the features \(X_1\) to \(X_4\) mentioned in the previous section as its input. As illustrated in Figure \ref{whole}, for these features, we subsequently input them into the Reverse Transformer Attention module for feature extraction.

\begin{figure}
  \centering
  \includegraphics[width=1.0\linewidth]{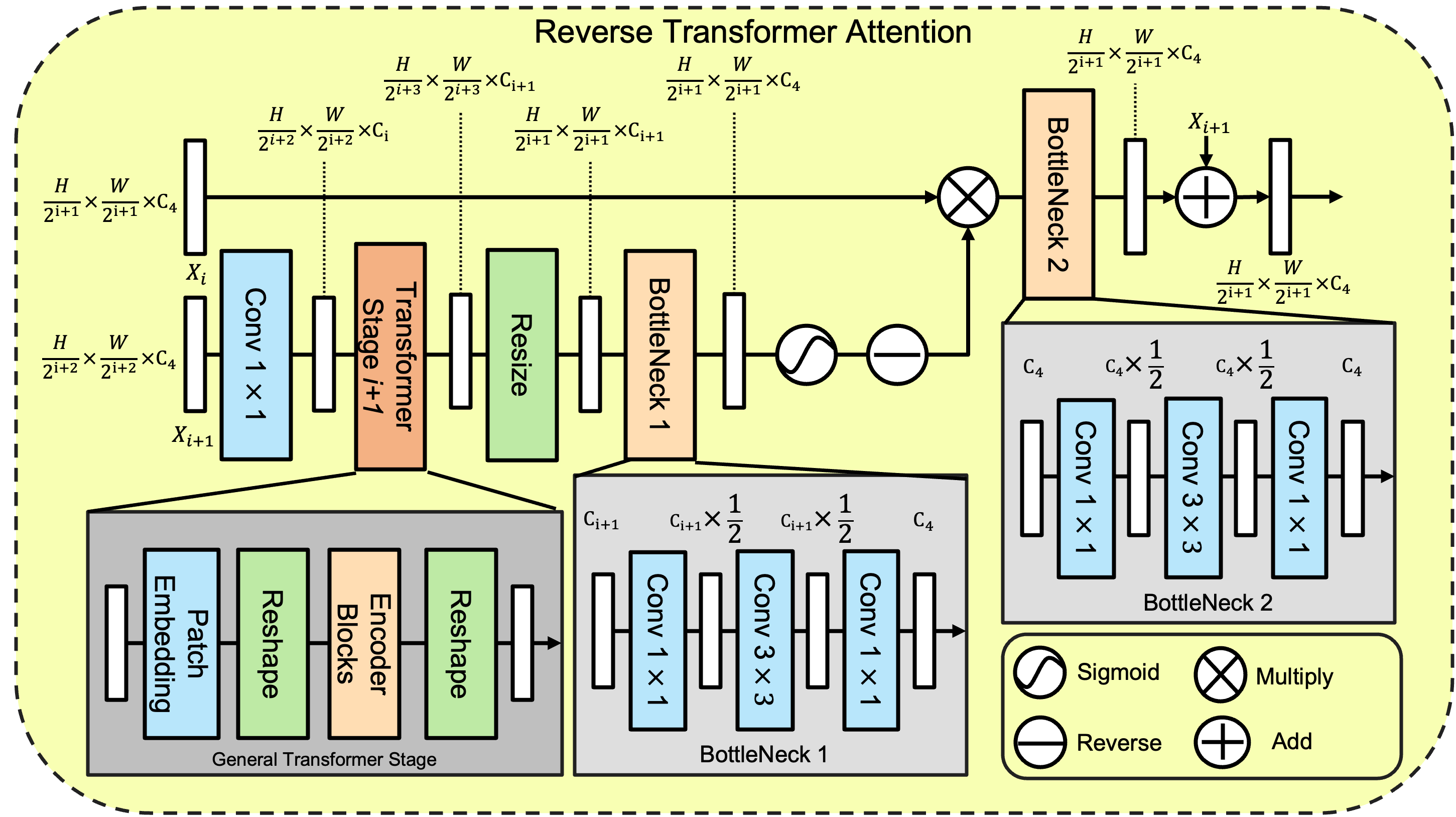}
  \caption{Structure of Reverse Transformer Attention (RTA)}
  \label{rta}
\end{figure}

\subsection{Reverse Transformer Attention (RTA)}
As illustrated in Figure \ref{rta}, the Reverse Transformer Attention (RTA) module employs a reverse attention mechanism for feature refinement. This module integrates multiple transformer stages for feature extraction, capitalizing on their ability to discern global and local features. This approach ensures that the significant but often neglected regions, especially the edge, are accentuated, enhancing the model's performance in capturing intricate details.

\textbf{Feature Processing:}
In the processing of sequential input features, $X_i$ and $X_{i+1}$, to align their dimensions with the transformer stage input, a convolutional layer, along with the transformer stage of the backbone, is employed. To maintain spatial consistency, features are resized and subsequently passed through a bottleneck structure to increase the dimensionality. The reverse attention mechanism is employed to emphasize typically overlooked regions by subtracting the attention map from the unity, which is formulated as follows:

\begin{equation}
    {X}_{i+1,\mathrm{reverse}} = 1 - \mathrm{BN}_1(\mathrm{Resize}(\mathrm{Stage}_{i+1}(\mathrm{Conv}({X}_{i+1})))).
\end{equation}

\textbf{Feature Enhancement and Output:}
The primary feature \( {X}_i \) is modulated using the reverse attention map, followed by refinement with a bottleneck structure:

\begin{equation}
    X_{i,output} = \mathrm{BottleNeck}_2({X}_i \odot {X}_{i+1,\mathrm{reverse}}) + {X}_{i}
\end{equation}

The output feature map \(X_{i,output}\) encapsulates enhanced information due to the reverse attention mechanism, enabling RTA-Former to dynamically emphasize different feature map regions.

\subsection{Decoder}

Starting with inputs \(X_1\) to \(X_4\) and four outputs \(X_{1,output}\) to \(X_{4,output}\) from the HFS, we perform the following operation:

\begin{align}
\vspace{-3em}
    &{X}_{\mathrm{fused},i} = \mathrm{Resize}(\mathrm{FF}(X_i, X_{i,output})),\\
    &M = \mathrm{Up}(\mathrm{Conv}(\mathrm{Cat}({X}_{\mathrm{fused},1}, {X}_{\mathrm{fused},2}, \ldots))).
    \vspace{-1.5em}
\end{align}

where \( {X}_{\mathrm{fused},i} \) is the result of feature fusion for each \( i \). \(\mathrm{FF}\) is the proposed feature fusion mechanism. For the final output \(M\), these features are resized, concatenated, passed through a convolutional layer, and then upsampled.

\section{EXPERIMENTS}
\subsection{Dataset}
We used five datasets for our experiments: CVC-ClinicDB\cite{bernal2017comparative}, CVC-ColonDB\cite{bernal2012cvc}, CVC-300\cite{vazquez2017benchmark}, ETIS-LaribPolypDB\cite{yang2022automatic}, and Kvasir\cite{jha2020kvasir}. These datasets provide a diverse and representative sample of gastrointestinal polyp images for developing and evaluating medical image segmentation models. Following the previous methods\cite{dong2021polyp,fan2020pranet}, we trained our models on a merged set comprising 550 images from CVC-ClinicDB and 900 from Kvasir, totaling 1,450 images. For testing, we evaluated the model's performance on all five datasets to assess both the model's learning capabilities and its generalization. 

\begin{table}[htbp]
\centering
\small
\caption{The param number of our four sizes of the model.}
\renewcommand{\arraystretch}{0.8} 
\begin{tabular}{ccc}
\hline
Model Name & Backbone & Param (M) \\
\hline
RTA-Former-T & PVTv2-B0 & 8.4 \\
RTA-Former-S & PVTv2-B2 & 56.2 \\
RTA-Former-M & PVTv2-B4 & 192.6 \\
RTA-Former-L & PVTv2-B5 & 250.8 \\
\hline
\end{tabular}
\label{model_params}
\end{table}

\subsection{Evaluation Metrics}
To quantitatively assess the network's performance on polyp segmentation, we used the Dice Similarity Coefficient (DICE) and mean Intersection over Union (mIoU).

\begin{table*}[htbp]
\small
\centering
\caption{Comparison Results of the purposed method on the 5 polyp segmentation datasets. \textcolor{blue}{Blue} indicates the best result, and \textcolor{red}{red} displays the second-best.}
\renewcommand{\arraystretch}{0.8} 
\begin{tabular}{ccccccccccc}
\toprule
Model & \multicolumn{2}{c}{Kvasir-SEG} & \multicolumn{2}{c}{CVC-ClinicDB} & \multicolumn{2}{c}{CVC-300} & \multicolumn{2}{c}{CVC-ColonDB} & \multicolumn{2}{c}{ETIS} \\
& DICE & mIoU & DICE & mIoU & DICE & mIoU & DICE & mIoU & DICE & mIoU \\
\midrule
MICCAI'15 U-Net\cite{ronneberger2015u} & 0.818 & 0.746 & 0.823 & 0.755 & 0.710 & 0.627 & 0.512 & 0.444 & 0.398 & 0.335 \\
DLMIA'18 UNet++\cite{zhou2018unet++} & 0.821 & 0.743 & 0.794 & 0.729 & 0.707 & 0.624 & 0.483 & 0.410 & 0.401 & 0.344 \\
MICCAI'20 ACSNet\cite{zhang2020adaptive} & 0.898 & 0.838 & 0.882 & 0.826 & 0.856 & 0.788 & 0.716 & 0.649 & 0.578 & 0.509 \\
arXiv'21 MSEG\cite{lambert2020mseg} & 0.897 & 0.839 & 0.909 & 0.864 & 0.874 & 0.804 & 0.735 & 0.666 & 0.700 & 0.630 \\
arXiv'21 DCRNet\cite{qin2020dcr} & 0.886 & 0.825 & 0.896 & 0.844 & 0.863 & 0.787 & 0.704 & 0.631 & 0.556 & 0.496 \\
MICCAI'20 PraNet\cite{fan2020pranet} & 0.898 & 0.840 & 0.899 & 0.849 & 0.871 & 0.797 & 0.712 & 0.640 & 0.628 & 0.567 \\
CRV'21 EU-Net\cite{patel2021enhanced} & 0.908 & 0.854 & 0.902 & 0.846 & 0.837 & 0.765 & 0.756 & 0.681 & 0.687 & 0.609 \\
MICCAI'21 SANet\cite{wei2021shallow} & 0.904 & 0.847 & 0.916 & 0.859 & 0.888 & 0.815 & 0.753 & 0.670 & 0.750 & 0.654 \\
arXiv'21 Polyp-PVT\cite{dong2021polyp} & 0.917 & 0.864 & 0.937 & \textcolor{red}{0.889} & 0.900 & 0.833 & \textcolor{red}{0.808} &\textcolor{red}{0.727} & 0.787 & 0.706 \\
IEEE TIM'23 APCNet\cite{yue2023attention} &0.913& 0.859 & 0.934 & 0.886 & 0.893 & 0.827 & 0.758 & 0.682&0.726 & 0.648 \\
PR'23 CFANet\cite{zhou2023cross} &0.915& 0.861 & 0.933 & 0.883 & 0.893 & 0.827 & 0.743 & 0.665&0.732 & 0.655 \\
SPIE'23 CaraNet\cite{lou2022CaraNet} & 0.918 & 0.865 & 0.936 & 0.887 & \textcolor{blue}{0.903} & \textcolor{blue}{0.838} & 0.773 & 0.689 & 0.747 & 0.672 \\
\hline

\rowcolor[HTML]{C8FFFD}  
RTA-Former-T (Ours) & 0.903 & 0.846 & 0.925 & 0.868 & 0.863 & 0.782 & 0.766 & 0.676 & 0.724 & 0.639 \\
\rowcolor[HTML]{C8FFFD}  
RTA-Former-S (Ours) & 0.920 & 0.866 & 0.931 & 0.883 & 0.893 & 0.822 & 0.794 & 0.711 & \textcolor{red}{0.789} & 0.710 \\
\rowcolor[HTML]{C8FFFD}  
RTA-Former-M (Ours) & \textcolor{red}{0.921} & \textcolor{red}{0.873} & \textcolor{blue}{0.939} & \textcolor{blue}{0.892} & \textcolor{red}{0.902} & \textcolor{red}{0.832} & 0.798 & 0.719 & \textcolor{red}{0.789} & \textcolor{red}{0.712} \\
\rowcolor[HTML]{C8FFFD} 
RTA-Former-L (Ours) & \textcolor{blue}{0.923} & \textcolor{blue}{0.875} & \textcolor{red}{0.938} & 0.888 & 0.891 & 0.815 & \textcolor{blue}{0.818} & \textcolor{blue}{0.734} & \textcolor{blue}{0.795} & \textcolor{blue}{0.714} \\
\bottomrule
\label{result1}
\end{tabular}
\end{table*}

\begin{table*}[htbp]
\small
\centering

\caption{Ablation Study of the purposed modules}
\begin{tabular}{lllllllllllll}
\hline
\multicolumn{3}{c}{Components} & \multicolumn{2}{c}{Kvasir-SEG} & \multicolumn{2}{c}{CVC-ClinicDB} & \multicolumn{2}{c}{CVC-300} & \multicolumn{2}{c}{CVC-ColonDB} & \multicolumn{2}{c}{ETIS} \\
HFS & RA  & RTA & DICE & mIoU & DICE & mIoU & DICE & mIoU & DICE & mIoU & DICE & mIoU \\
\hline
       &   &                & 0.909        & 0.855       & 0.906           & 0.851          & 0.875       & 0.799       & 0.804          & 0.72           & 0.763       & 0.683      \\
\ding{51}       &   &                 & 0.914        & 0.861       & 0.912           & 0.859          & 0.889       & 0.806       & 0.806          & 0.723          & 0.779       & 0.696      \\
\ding{51}       & \ding{51} &                 & 0.915        & 0.863       & 0.922           & 0.869          & 0.890        & 0.814       & 0.811          & 0.726          & 0.788       & 0.704      \\
\ding{51}       &  &  \ding{51}              & \textbf{0.923}        & \textbf{0.875}       & \textbf{0.938}           & \textbf{0.888}          & \textbf{0.891}       & \textbf{0.815}       & \textbf{0.818 }         & \textbf{0.734}          & \textbf{0.795}       & \textbf{0.714}      \\ \hline
\label{ablation}
\end{tabular}
\end{table*}

\subsection{Implementation}
All models are trained on a cluster with 8 NVIDIA RTX 6000 GPUs, each with 24GB memory, utilizing CUDA 12.2. We set the learning rate at \(1 \times 10^{-4}\) and the weight decay rate at \(1 \times 10^{-4}\), batch size 8, Adam optimizer across 100 epochs with structure loss\cite{fan2020pranet}. 

 To ensure a fair evaluation of our method, we adhere to the image resolution settings previously used in each dataset. For the polyp segmentation task, we resize images to 352x352 as previous methods\cite{dong2021polyp, fan2020pranet}, with scales {0.75, 1.0, 1.25}. 

In Table \ref{model_params}, we present four distinct sizes of our model: tiny (T), small (S), medium (M), and large (L). These variations stem from our goal to understand how models with different numbers of parameters fare across diverse tasks. Balancing complexity becomes essential to prevent potential overfitting while ensuring the model effectively grasps medical image intricacies. These four models, T, S, M, and L, are respectively built upon the encoder backbone PolyPVT-v0, v2, v4, and v5, allowing users to adapt to various scenarios and computational limits.

\subsection{Results}
In Table \ref{result1}, we demonstrated the result based on the DICE and the mIoU metrics. The Medium and Large RTA-Former models stand out in their learning capabilities, surpassing several models on Kvasir-SEG and CVC-ClinicDB with DICE scores of up to 93.9\% and mIoU scores of up to 89.2\%. PVT-based methods, notably RTA-Former, consistently outperformed many CNN-based approaches such as UNet and UNet++. Moreover, RTA-Former's performance on CVC-300, CVC-ColonDB, and ETIS datasets underscores its impressive generalization. While some networks like the Medium model and CaraNet excel on smaller datasets like CVC-300, many CNN-based models, particularly UNet and UNet++, lagged in generalization. Our RTA-Former effectively captures and analyzes polyp representations, outperforming the other 12 models in both learning and generalization, especially for RTA-Former-M and RTA-Former-L.

\subsection{Ablation Study}

\textbf{Impacts of Our Modules.} As shown in Table \ref{ablation}, we used PVTv2-B5 as our base model. Subsequently, we incorporated the hierarchical feature synthesizer (HFS) without the RTA module. Building upon the HFS, we introduced the traditional convolution-layered reverse attention (RA) which is introduced in CaraNet\cite{lou2022CaraNet}. In the final step, we integrated our proposed reverse transformer attention (RTA) with the HFS. It shows that the introduction of HFS yields approximately a 1\% enhancement in both DICE and mIoU metrics across all datasets. Employing solely the RA exhibits marginal improvement for Kvasir-SEG and CVC-300, while other datasets observe a proximate 1\% increment. Contrarily, with the RTA's integration, underpinned by the transformer, a notable 2\% augmentation is observed on several datasets, thereby accentuating RTA's adaptability and potency over RA.

\begin{figure}[htbp]
  \centering
  \includegraphics[width=1.0\linewidth]{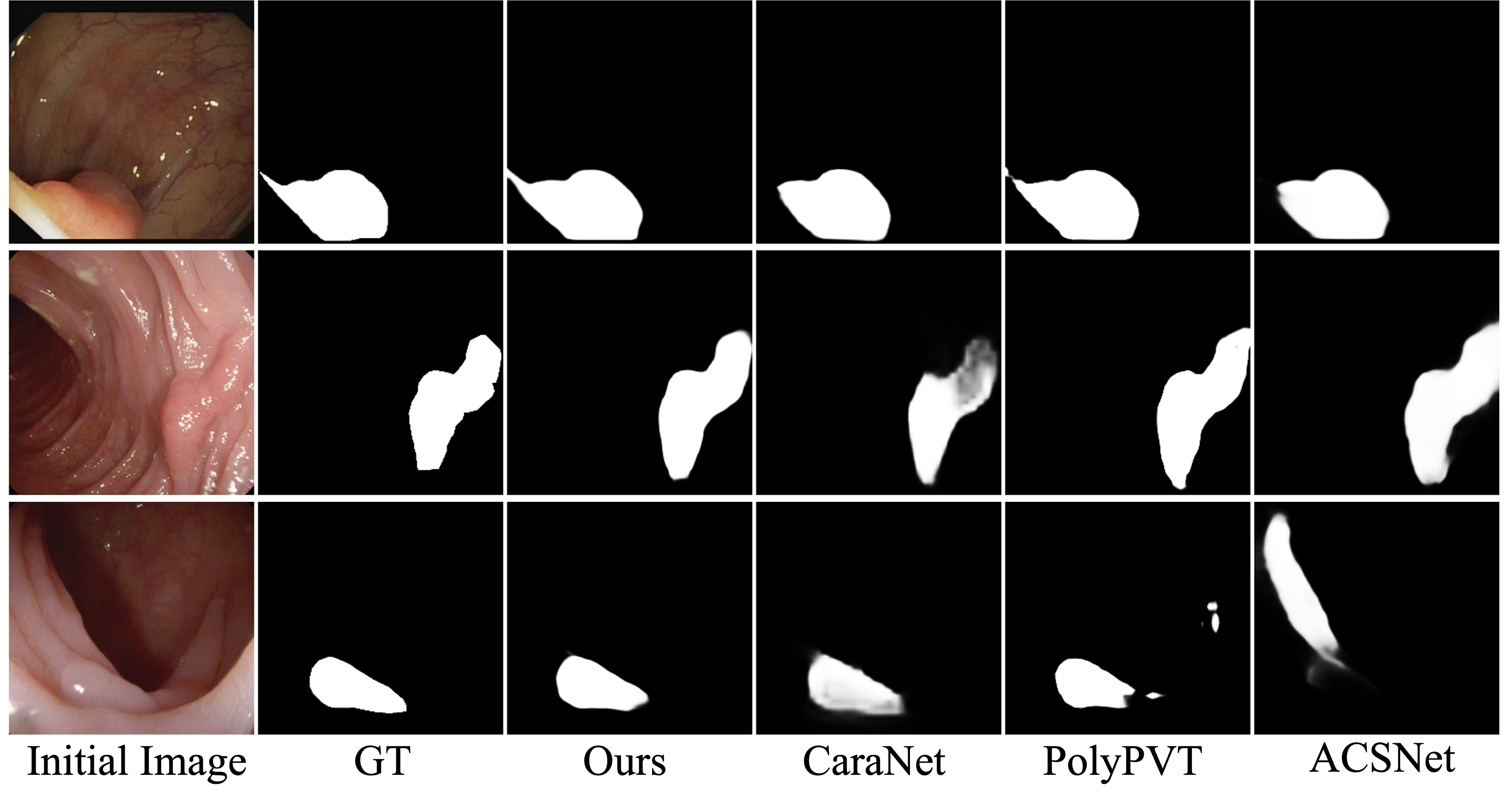}
\caption{Visualization comparison of polyp segmentation results for our model and other models on polyps of varying scales. GT refers to the ground truth of the dataset annotations. The last four columns show the prediction masks generated by the models.}
  \label{vis1_poly}
\end{figure}

\begin{figure}[htbp]
  \centering
  \includegraphics[width=1.0\linewidth]{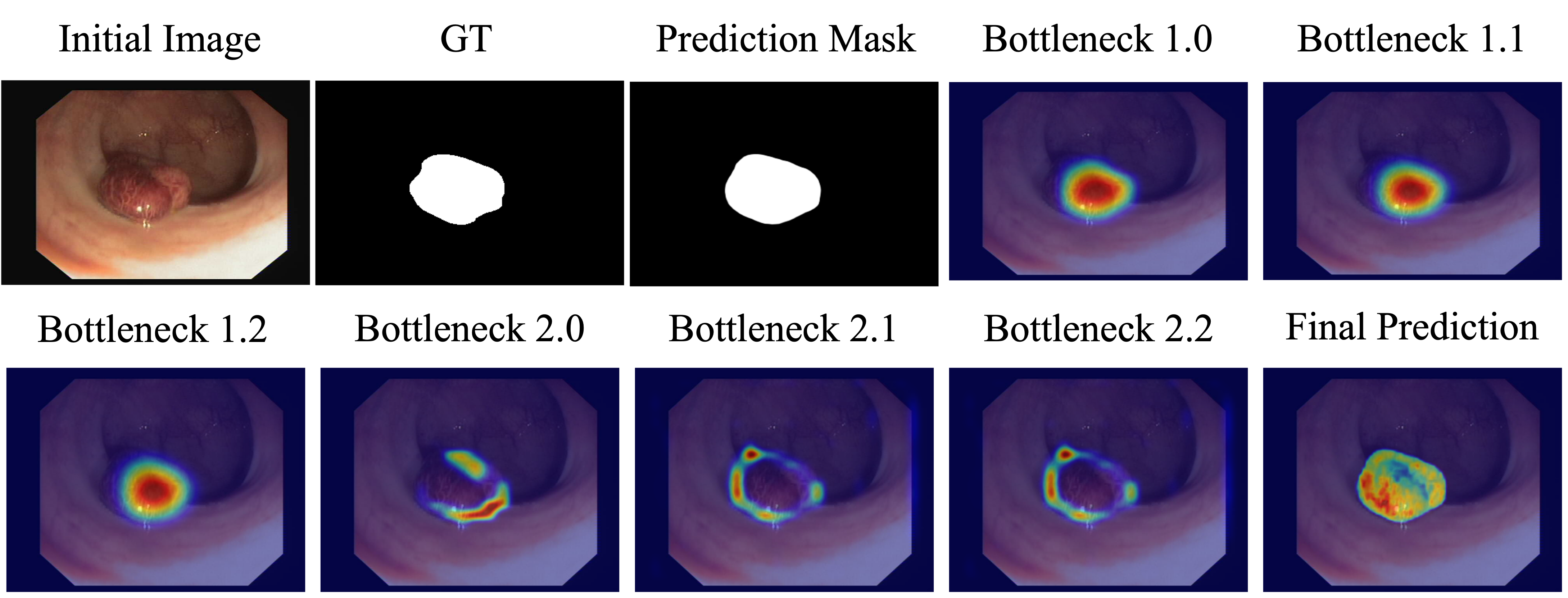}
  \caption{Visualization of our attention module. Bottleneck 1.0 to Bottleneck 1.2 are the feature maps before the reverse mechanism. Bottleneck 2.0 to Bottleneck 2.2 are the feature maps after the reverse mechanism. }
  \label{vis2_poly}
\end{figure}

\subsection{The Visualization Result of the Polyp Segmentation.}

In Figure \ref{vis1_poly}, we show the sensitivity and segmentation performance of different models towards polyps of varied scales on the test dataset from ClinicDB and ETIS. From the results, it becomes evident that our model consistently demonstrates robust segmentation capabilities across polyps of all sizes, outperforming other models in its adaptability and precision. Additionally, our model exhibits superior sensitivity towards the ambiguous or blurry edges of the polyps, ensuring more accurate segmentation even in challenging scenarios, which will be discussed next.

\subsection{The Visualization Result of Our Reverse Attention Module.}

We employ the Grad-CAM\cite{selvaraju2017grad} technique to illustrate the region of focus within our model. As shown in Figure \ref{rta}, the output from Bottleneck 1 is reversed and subsequently fed into Bottleneck 2. As shown in Figure \ref{vis2_poly}, within each bottleneck structure, Bottleneck 1.0, 1.1, and 1.2 denote the three convolution layers that sequentially process the image features. Similarly, Bottleneck 2.0, 2.1, and 2.2 correspond to the convolution layers within the second bottleneck unit. This process reveals that the original output of Bottleneck 1 is concentrated on the polyp itself, whereas the inverted feature map predominantly highlights the periphery of the polyp region. This distinction underscores the efficacy of our methodology in accurately delineating the lesion's boundary, thereby facilitating enhanced segmentation.

\section{CONCLUSIONS}
This study introduces a novel approach termed RTA-Former, wherein the encoder employs PVT and the decoder incorporates the RTA module to enrich the reverse attention mechanism. The model shows powerful capability and generalization on various datasets. The network can employ different versions of transformers according to the complexity of the task. In particular, RTA-Former-L and RTA-Former-M exhibit the highest performance levels across datasets such as Kvasir-SEG, CVC-ClinicDB, CVC-ColonDB, and ETIS in the context of polyp segmentation. These are promising applications of RTA-Former in polyp segmentation.







{\small
\bibliographystyle{IEEEbib}
\bibliography{egbib.bib}
}

\end{document}